\documentclass[a4paper,12pt]{article}

\usepackage{amsmath}
\usepackage{amssymb,epsfig,psfig}

\newtheorem{lemma}{Lemma}[section]

\newcommand{\be}{\begin{equation}}
\newcommand{\ee}{\end{equation}}
\newcommand{\bea}{\begin{eqnarray}}
\newcommand{\eea}{\end{eqnarray}}
\newcommand{\sh}{\sinh}

\newcommand{\prf}{\noindent{\bf Proof}\ }
\newcommand{\qed}{{\hfill $\Box$}}

\begin{document}

\title{Parametric Representation\\ of Noncommutative Field Theory}
\author{Razvan Gurau and Vincent Rivasseau\footnote{e-mail: Razvan.Gurau@th.u-psud.fr; Vincent.Rivasseau@th.u-psud.fr}\\
Laboratoire de Physique Th\'eorique, CNRS UMR 8627\\
Universit\'e Paris XI\footnote{Work supported by ANR grant NT05-3-43374 ``GenoPhy".}}
\maketitle

\begin{abstract}
In this paper we investigate the Schwinger parametric
representation for the Feynman amplitudes of the recently discovered
renormalizable $\phi^4_4$ quantum field theory on the Moyal non commutative
${\mathbb R^4}$ space. This representation involves new {\it hyperbolic} polynomials
which are the non-commutative analogs of the usual ``Kirchoff"
or ``Symanzik" polynomials of commutative field theory, 
but contain richer topological information.
\end{abstract}

\section{Introduction}

Non-commutative field theories (for a general review see \cite{DN}) deserve a thorough 
and systematic investigation. Indeed they may be relevant for physics beyond the standard model.
They are certainly effective models for certain limits of string theory
\cite{CDS}-\cite{SW}. What is often less emphasized is that they
can also describe effective physics in our ordinary standard world but
with non-local interactions, such as the physics of the quantum Hall effect
\cite{Su}.

In joint work with J. Magnen and F. Vignes-Tourneret \cite{GMRV},
we provided recently a new proof that the Grosse-Wulkenhaar
scalar $\Phi^4$ theory on the Moyal space ${\mathbb R}^4$, hereafter 
called NC$\Phi^4_4$, is renormalizable to all orders in perturbation theory using direct space multiscale analysis.

The Grosse-Wulkenhaar breakthrough \cite{GW1}\cite{GW2}
found that the right propagator in non-commutative field theory 
is not the ordinary commutative propagator, but has to be modified to obey Langmann-Szabo duality \cite{LS},\cite{GW2}. Grosse and Wulkenhaar added an harmonic potential which
can be interpreted as a piece of the covariant Laplacian in a constant magnetic field. They computed the
corresponding ``vulcanized" propagator in the ``matrix base" which transforms the Moyal product 
into a matrix product. They use this representation to prove 
perturbative renormalizability of the theory up to some estimates 
which were finally proven in \cite{RVW}.

Our direct space method builds upon the previous work of Filk \cite{F} 
who introduced clever simplifications,
also called ``Filk moves", to treat the combination of 
oscillations and $\delta$ functions which characterize non commutative interactions. 
Minwalla, van Raamsdonk and Seiberg \cite{MvRS} computed a Schwinger parametric 
representation for the ``not-vulcanized" $\Phi^4_4$ non-commutative theory.
Subsequently Chepelev and Roiban computed also such a Schwinger parametric 
representation for this theory in \cite{CR1}
and used it in \cite{CR2} to analyze power counting. 
These works however remained inconclusive, since they 
worked with the vertex but not the right propagator of NC$\Phi^4_4$, where
ultraviolet/infrared mixing prevents from obtaining a finite 
renormalized perturbation series. We have also been unable to find up to now 
the proofs for the formulas in \cite{MvRS}-\cite{CR1}, which in fact disagree.

The parametric representation introduced in this work is completely 
different from the ones of \cite{CR1} or \cite{MvRS},
since it corresponds to the renormalizable vulcanized theory.
It no longer involves direct polynomials in the Schwinger parameters
but new polynomials of hyperbolic functions of these
Schwinger parameters. This is because the propagator of NC$\Phi^4_4$ is based on
the Mehler kernel rather than on the ordinary heat kernel.
These hyperbolic polynomials contain richer topological information
than in ordinary commutative field theory. Based on ribbon graphs, they 
contain information about their invariants, such
as the genus of the surface on which these graphs live.

This new parametric representation is a compact tool for the study of non commutative field theory which has the advantages
(positivity, exact power counting) but not the drawbacks 
(awkwardness of the propagator) of the matrix base representation.
It can be used as a starting point to work out 
the renormalization of the model directly in parametric space, as can be done
in the commutative case \cite{BL}. It is also a good starting point
to define the regularization and minimal dimensional
renormalization scheme of NC$\Phi^4_4$. This  dimensional scheme in the ordinary field theory case better preserves continuous symmetries such as gauge symmetries, hence played a historic
role in the proof of `tHooft and Veltman that non Abelian gauge theories on commutative ${\mathbb R^4}$ are renormalizable. It is also 
used extensively in the works of Kreimer and Connes \cite{K}\cite{CK} which recast 
the recursive BPHZ forest formula of perturbative renormalization into 
a Hopf algebra structure and relate it to a new class of Riemann-Hilbert problems; 
here the motivations to use dimensional renormalization
rather than e.g. subtraction at zero momentum come 
at least in part from number theory rather than from physics.

Following these works, renormalizability has also attracted considerable 
interest in the recent years as a pure mathematical structure. 
The renormalization group ``ambiguity" reminds mathematicians of
the Galois group ambiguity for roots of algebraic equations \cite{CoM}.
Finding new renormalizable theories may therefore be 
important for the future of pure mathematics as well as for physics.

This paper is organized as follows. In Section II we introduce notations
and define our new polynomials $HU$ and $HV$ which generalize the Symanzik
polynomials $U$ and $V$ of commutative field theory. In section III 
we prove the basic positivity property of the first polynomial $HU$
and compute leading ultraviolet terms which allow to recover 
the right power counting in the parametric 
representation, introducing a technical trick which we call
the ``third Filk move"
\footnote{For technical reasons
exact power counting was not fully established in \cite{GMRV}.}.
Section IV establishes the positivity properties 
and computes such leading terms for the second polynomial
$HV$, the one which gives the dependence in the external arguments.
Finally examples of these polynomials for various graphs 
are given in section V.


\section{Hyperbolic Polynomials}
\subsection{Notations}

The NC$\Phi^4_4$ theory is defined on ${\mathbb R}^4$ equipped
with the associative and noncommutative Moyal product
\begin{align}
(a\star b)(x) &= \int \frac{d^4k}{(2\pi)^4} \int d^4 y \; a(x{+}\tfrac{1}{2}
\theta {\cdot} k)\, b(x{+}y)\, \mathrm{e}^{\mathrm{i} k \cdot y}\;.
\label{starprod}
\end{align}

The renormalizable action functional introduced in \cite{GW2} is
\begin{equation}\label{action}
S[\phi] = \int d^4x \Big( \frac{1}{2} \partial_\mu \phi
\star \partial^\mu \phi + \frac{\Omega^2}{2} (\tilde{x}_\mu \phi )
\star (\tilde{x}^\mu \phi ) + \frac{1}{2} \mu_0^2
\,\phi \star \phi 
+ \frac{\lambda}{4!} \phi \star \phi \star \phi \star
\phi\Big)(x)\;,
\end{equation}
where the Euclidean
metric is used. In what follows the mass $\mu_0$ does not play any role so we put it to zero\footnote{This does not lead in this model to any infrared divergences.
Beware that our definition of $\Omega$ is different from the one of \cite{GW2}
by a factor $4\theta^{-1}$}.

In four dimensional $x$-space the propagator is \cite{GRV}
\begin{equation}\label{tanhyp}
\int\prod_l \frac{\Omega d\alpha_l}{[2\pi\sinh(\alpha_l)]^{D/2}}
e^{-\frac{\Omega}{4}\coth(\frac{\alpha_l}{2})u_l^2-
\frac{\Omega}{4}\tanh(\frac{\alpha_l}{2})v_l^2}\; .
\end{equation} 
and the (cyclically invariant) vertex is:
\begin{equation}\label{vertex}
V(x_1, x_2, x_3, x_4) = \delta(x_1 -x_2+x_3-x_4 )e^{i
\sum_{1 \le i<j \le 4}(-1)^{i+j+1}x_i \theta^{-1}  x_j}\ ,
\end{equation}
where we note $x \theta^{-1}  y  \equiv  \frac{2}{\theta} (x_1  y_2 -  x_2  y_1 +  
x_3  y_4 - x_4  y_3 )$.

Permutational symmetry of the fields at all vertices, which characterizes commutative field theory,
is replaced by the more restricted cyclic symmetry. Hence
the ordinary Feynman graphs of $\Phi^4_4$ really become ribbon 
graphs in NC$\Phi^4_4$. For such a ribbon 
graph $G$, we call $n$, $L$, $N$, $F$, and $B$
respectively the number of vertices, of internal lines, of external half-lines, of faces and of
faces broken by some external half-lines. The Euler characteristic
is $2-2g = n-L+F$, where $g$ is the genus of the graph.
To each graph $G$ is associated a dual graph ${\cal G}$ of same genus by
exchanging faces and vertices.

In ordinary commutative field theory, in order to obtain Symanzik's polynomials it is not convenient
to solve the momentum conservation at the vertices through a momentum routing, because
this is not canonical. It is better to express 
these $\delta$ functions through their Fourier transform. After integration
over internal variables, the amplitude of an 
amputated graph $G$ with external momenta $p$ is, up to a normalization,
in space time dimension $D$ (of course the main case of interest 
in this paper is $D=4$):
\bea \label{symanzik} 
A_G (p) = \delta(\sum p)\int_0^{\infty} 
\frac{e^{- V_G(p,\alpha)/U_G (\alpha) }}{U_G (\alpha)^{D/2}} 
\prod_l  ( e^{-m^2 \alpha_l} d\alpha_l )\ .
\eea
The first and second Symanzik polynomials $U_G$ and $V_G$ are
\begin{equation}\label{symanzik1}
U_G = \sum_T \prod_{l \not \in T} \alpha_l \ ,
\end{equation}
\begin{equation}\label{symanzik2}
V_G = \sum_{T_2} \prod_{l \not \in T_2} \alpha_l  (\sum_{i \in E(T_2)} p_i)^2 \ , 
\end{equation}
where the first sum is over spanning trees $T$ of $G$
and the second sum  is over two trees $T_2$, i.e. forests separating the graph
in exactly two connected components $E(T_2)$ and $F(T_2)$; the corresponding
Euclidean invariant $ (\sum_{i \in E(T_2)} p_i)^2$ is, by momentum conservation, also
equal to $ (\sum_{i \in F(T_2)} p_i)^2$.

The topological formulas (\ref{symanzik1}) and (\ref{symanzik2}) are
a field-theoretic instance of the tree-matrix theorem of Kirchoff et al; 
for a recent review of this kind of theorems see \cite{A}.

In the non commutative case,
momentum routing is replaced by position routing \cite{GMRV}.
However this position routing 
is again non-canonical, depending on the choice of a particular tree. 
Therefore to compute the parametric representation we prefer to perform
a new level of Fourier transform: we represent the ``position conservation" 
rules as integrals
over new ``hypermomenta" $p_v$ associated to each of the vertices:
\begin{equation}
\delta(x_1 -x_2+x_3-x_4 ) = \int  \frac{d p_v}{(2 \pi)^D}
e^{ip_v(x_1^v-x_2^v+x_3^v-x_4^v)} = \int  \frac{d p_v}{(2 \pi)^D}
e^{p_v \sigma (x_1^v-x_2^v+x_3^v-x_4^v)}
\end{equation}
where $\sigma $ is the $D$ by $D$ matrix defined by $D/2$ matrices $\sigma_y$
on the diagonal (we assume $D$ even):
\bea
\sigma=\begin{pmatrix}
\sigma_y & \cdots & 0 \\
\cdots & \cdots & \cdots \\
0 & \cdots & \sigma_y
\end{pmatrix}  \ \ {\rm where} \ \ \sigma_y=\begin{pmatrix}
0 & -i \\
i & 0 \end{pmatrix} \ .
\eea

There is here a subtle difference with the commutative case.
The first commutative polynomial (\ref{symanzik1}) is not the determinant of the quadratic
form in the internal position variables, since this determinant vanishes by translation invariance.
It is rather the determinant of the quadratic form integrated over all internal positions of 
the graph {\it save one} (remark the overall 
momentum conservation in (\ref{symanzik})). This is a canonical object which does
not depend of the choice of the particular vertex whose position is not integrated 
(this can be seen explicitly on
the form (\ref{symanzik1}), which depends only on $G$).

In the non commutative case translation invariance is lost. This allows to define the amplitude of a graph
as a function of the external positions by integrating over all internal positions 
and hypermomenta, since the corresponding 
determinant no longer vanishes. In this way one can 
define {\it canonical} polynomials $HU_G$ and $HV_G$
which only depend on the ribbon graph $G$. 

But in practice it is often more convenient (for instance for renormalization or for understanding the
limit towards the commutative case)  to define the amplitude of a graph 
by integrating all the internal positions and hypermomenta save one, $p_{\bar{v}}$;
this helps to factorize an overall approximate ``position conservation" for the whole graph. 
However precisely because there is no translation invariance, the corresponding
polynomials $HU_{G,\bar{v}}$ and $HV_{G,\bar{v}}$ explicitly depend
on the ``rooted graph" $G,\bar{v}$, i.e. on the choice of $\bar{v}$ 
(although their leading ultraviolet terms do not depend on this choice, see below).

Consider a graph $G$ with $n$ vertices, $N$ external positions 
and a set $L$ of $2n-N/2$ internal lines or propagators.
Each vertex in NC$\phi^4$ is made of four ``corners",
bearing either a halfline or an external field, 
numbered as $1,2,3,4$ in the cyclic order given by the Moyal product. 
To each such corner is associated a position, noted $x_i$. The set $I$ of internal corners 
has $4n - N$ elements, labeled usually as $i,j,...$; the set $E$ of 
external corners has $N$ elements labeled as $e,e',...$.
A line $l$ of the graph joins two corners in $I$, with positions $(x^l_i,x^l_j)$ 
(which in general do not belong to the same vertex).

The amplitude of such a NC$\phi^4$ graph $G$ is then given, up to some inessential normalization $K$, by:
\bea
{\cal A}_G(\{x_e\})&=& K
\int \prod_{l} \frac{d\alpha_l}{\sh\alpha_l^{D/2}} \int \prod_{i \in I}d x_i
\prod_{v} d p_v\nonumber\\
&&\prod_l e^{-\frac{\Omega}{4}\coth(\frac{\alpha_l}{2})(x_i^l-x_j^l)^2-
\frac{\Omega}{4}\tanh(\frac{\alpha_l}{2})(x_i^l+x_j^l)^2}\nonumber\\
&&\prod_v  e^{\frac{i}{2}\sum_{1\le i<j\le 4}(-1)^{i+j+1}x_i^v\theta^{-1} x_j^v+
p_v \sigma (x_1^v-x_2^v+x_3^v-x_4^v)}
\eea
or by 
\bea
{\cal A}_{G,{\bar v}}(\{x_e\},\; p_{\bar v})&=& K
\int \prod_l \frac{d\alpha_l}{\sh\alpha_l^{D/2}} \int \prod_{i \in I}d x_i
\prod_{v\neq\bar{v}} d p_v \nonumber\\
&&\prod_l e^{-\frac{\Omega}{4}\coth(\frac{\alpha_l}{2})(x_i^l-x_j^l)^2-
\frac{\Omega}{4}\tanh(\frac{\alpha_l}{2})(x_i^l+x_j^l)^2}\nonumber\\
&&\prod_v  e^{\frac{i}{2}\sum_{1\le i<j\le 4}(-1)^{i+j+1}x_i^v\theta^{-1} x_j^v+p_v \sigma (x_1^v-x_2^v+x_3^v-x_4^v)}\ .
\eea

\subsection{Definition of $HU$ and $HV$}
The fundamental observation is that the integrals to perform being Gaussian, the result is
a Gaussian in the external variables divided by a determinant. This gives the definition of 
our hyperbolic parametric representation.
We introduce the notations $c_l = \coth(\frac{\alpha_l}{2}) = 1/t_l $
and $t_l = \tanh(\frac{\alpha_l}{2}) $. 
Using $\sinh \alpha_l = 2 t_l  / (1-t_l^2)  $ we obtain
\bea\label{hyperpcan}
{\cal A}_G (\{x_e\}) = K  \int_{0}^{\infty} \prod_l  [ d\alpha_l (1-t_l^2)^{D/2} ]
HU_G ( t )^{-D/2}   e^{-  \frac {HV_G ( t , x_e )}{HU_G ( t )}},
\eea
\bea\label{hyperpnoncan}
{\cal A}_{G,{\bar v}}  (\{x_e\},\;  p_{\bar v}) = K'  \int_{0}^{\infty} \prod_l  [ d\alpha_l (1-t_l^2)^{D/2} ]
HU_{G, \bar{v}} ( t )^{-D/2}   
e^{-  \frac {HV_{G, \bar{v}} ( t , x_e , p_{\bar v})}{HU_{G, \bar{v}} ( t )}},
\eea
where $K$ and $K'$ are some new inessential normalization constants
(which absorb in particular the factors 2 from $\sinh \alpha_l = 2 t_l  / (1-t_l^2)  $); $HU_G ( t )$ or $HU_{G, \bar{v}} ( t )$
are polynomials in the $t$ variables (there are no $c$'s because they are compensated
by the $t$'s coming from
$\sinh \alpha_l = 2 t_l  / (1-t_l^2)  $)
and $HV_G ( t , x_e )$ or $HV_{G, \bar{v}} ( t , x_e , p_{\bar v})$ are
quadratic forms in the external variables $x_e$ or  $(x_e , p_{\bar v})$ whose coefficients
are polynomials in the $t$ variables (again there are no $c$'s because they are compensated
by the $t$'s  which were included in the definition of $HU$, see below the difference between (\ref{HVgv1}) and (\ref{HVgv})).

There is a subtlety here. Overall approximate ``position conservation"
holds only for orientable graphs in the sense of \cite{GMRV}. 
Hyperbolic polynomials  for non orientable graphs  are
well defined through formulas  (\ref{hyperpcan})-(\ref{hyperpnoncan}) 
but they are significantly harder to compute than in the orientable case. 
Their amplitudes are also smaller in the ultraviolet,  and in
particular do not require any renormalization. Also many interesting 
non commutative theories such as the LSZ models \cite{LSZ}, the 
more general $(\bar\phi\phi)^2$ models of \cite{GMRV} and
the most natural Gross-Neveu models \cite{GRV}, \cite{V} do not have any 
non orientable subgraphs. So for simplicity we shall restrict ourselves 
in this paper to examples 
of hyperbolic polynomials for orientable graphs; and
when identifying leading pieces 
under global scaling in the hyperbolic polynomials, something necessary for renormalization, we also limit ourselves to the orientable case.

We now proceed to the computation of these hyperbolic polynomials.

\subsection{Short and Long Variables}

This terminology was introduced in \cite{GMRV}.

We order each line $l$ joining corners $l=(i,j)$ 
(which in general do not belong to the same vertex), 
in an arbitrary way such that it exits $i$ and enters $j$. 
We define the incidence matrix 
between lines and corners $\epsilon_{l i}$ to be $1$ if $l$ enters in $v$, $-1$ if it exits at $i$
and 0 otherwise. 
Also we define $\eta_{l i}=|\epsilon_{l i}|$. We note the property:
\bea
\sum_l(\epsilon_{l i}\epsilon_{l j}+\eta_{l i}\eta_{l j})=2\delta_{i j} \ .
\eea
We now define the short variables $u$ and the long variables $v$ as
\be
v_l=\sum_i\frac{\eta_{l i}x_i}{\sqrt 2},~u_l=\sum_i\frac{\epsilon_{l i} x_i}{\sqrt 2}
;~x_i=\sum_l \frac{\eta_{l i}v_l+\epsilon_{l i}u_l}{\sqrt{2}}\ .
\ee
The Jacobian of this change of coordinates is $1$.
Moreover, in order to avoid unpleasant $\sqrt{2}$ factors we rescale the external positions to hold $\bar{x}_e=\sqrt{2}x_e$ and the internal hypermomenta $\bar{p}_v=p_v/\sqrt{2}$.
Note that if the graph is orientable we can choose $\epsilon_{l i}$ 
to be $(-1)^{i+1}$, so that the incidence matrix is consistent with the 
cyclic order at the vertices (halflines alternatively enter and go out).
The integral in the new variables is:
\bea
&&\int \prod_l \big[\frac{1-t_l^2}{t_l}\big]^{D/2} d\alpha_l  \int \prod_{i}d x_i  \prod_{v\neq\bar{v}} d p_v
\prod_l e^{-\frac{\Omega}{2}\coth(\frac{\alpha_l}{2})u_l^2-
\frac{\Omega}{2}\tanh(\frac{\alpha_l}{2})v_l^2}\nonumber\\
&&  \prod_v  e^{\frac{i}{4}\sum_{\substack{i < j;\\ i,j \in v}}(-1)^{i+j+1}(\eta_{l i}v_l+\epsilon_{l i}u_l)\theta^{-1}
(\eta_{l' j}v_{l'}+\epsilon_{l' j}u_{l'})}
\prod_v e^{ \bar{p}_v \sigma \sum_{i\in v}(-1)^{i+1}(\eta_{l i}v_l+\epsilon_{l i}u_l)}
\nonumber\\
&&e^{\frac{i}{4}[\sum_{i \ne e}\omega(i,e)(\eta_{l i}v_l+\epsilon_{l i} u_l)
\theta^{-1}\bar{x}_e ]+\frac{i}{4}\sum_{e < e' } \bar{x}_e\theta^{-1}\bar{x}_{e'}+\sum_{e\in v}\bar{p}_v \sigma
(-1)^{e+1} \bar{x}_e}\ ,
\eea
where $\omega(i,e) = 1$ if $i<e$ and $\omega(i,e) = -1$ if $i>e$. When we write $i \in v$, it means that the corner $i$ belongs to $v$. 
>From now on we forget the bar over the rescaled variables.
We also concentrate on the computation of $HU_{G, {\bar v}}$ in (\ref{hyperpnoncan});
we indicate alongside the necessary modifications for $HU_{G}$ in (\ref{hyperpcan}).

We introduce the condensed notations:
\bea
{\cal A}_G  = \int \big[\frac{1-t^2}{t}\big]^{D/2} d\alpha \int d x d p e^{-\frac{\Omega}{2} X G X^t}
\eea
where 
\bea\label{defMPQ1}
X = \begin{pmatrix}
x_e & \bar{p} & u & v & p\\
\end{pmatrix} \ \ , \ \  G= \begin{pmatrix} M & P \\ P^{t} & Q \\
\end{pmatrix}\ .
\eea

Gaussian integration gives, up to inessential constants:
\bea\label{defMPQ2}
{\cal A}_G  = \int \big[\frac{1-t^2}{t}\big]^{D/2} d\alpha\frac{1}{\sqrt{Q}}
e^{- \frac{\Omega}{2}   
\begin{pmatrix} x_e & \bar{p} \\
\end{pmatrix} [M-P Q^{-1}P^{t}]
\begin{pmatrix} x_e \\ \bar{p} \\
\end{pmatrix}  }\ .
\eea

All we have to do now to get $HU$ and $HV$ is to compute the determinant and the minors of the matrix $Q$ for an arbitrary graph.

\section{The Determinant or First Hyperbolic Polynomial $HU$}

We define $I_D$ to be the identity matrix in $D$ dimensions. We also put $d=2L+n-1$ so that
$Q$ can then be written as:
\bea
Q=A\otimes I_{D} +B\otimes \sigma
\eea
with $A$ is a $d$ by $d$ symmetric matrix (accounting for the contribution of the propagators in 
the Gaussian) and $B$ a antisymmetric matrix (accounting for the oscillation part in the Gaussian).

Note that the  symplectic pairs decouple completely so that 
$\det Q=[\det(A\otimes I_2+B\otimes\sigma_y)]^{D/2}$. 

\begin{lemma} For any two $n\times n$ matrices 
$A$ and $B$ let $R=A\otimes I_2+B\otimes\sigma_y$. Then:
\bea \label{rdeterminant}
\det R=(-1)^n\det(A+B) \det(A-B)
\eea
and:
\bea\label{rinverse}
R^{-1}=\frac{[(A+B)^{-1}+(A-B)^{-1}]}{2}\otimes
I_2+\frac{[(A+B)^{-1}-(A-B)^{-1}]}{2}\otimes\sigma_y .
\eea
\end{lemma}

\prf We express the determinant as a Grassmann-Berezin integral:
\bea
&&\Delta=\det(A\otimes I_2+B\otimes \sigma_y)\nonumber\\
&=&\int \prod_k d\bar{\psi}^1_k d\psi^1_k
d\bar{\psi}^2_k d^2\psi_k
e^{-\begin{pmatrix}
\bar{\psi}^1_i & \bar{\psi}^{2}_i\\
\end{pmatrix}
(a_{i j}\otimes I_2+b_{i j}\otimes\sigma)
\begin{pmatrix}
\psi^{1}_{j} \\ \psi^{2}_{j} \\
\end{pmatrix}
}\nonumber\\
&=&\int \prod_k d\bar{\psi}^1_k d\psi^1_k
d\bar{\psi}^2_k d^2\psi_k
e^{-[a_{i j}(\bar{\psi}^1_i\psi^1_j+
\bar{\psi}^2_i\psi^2_j)
+ib_{i j}(-\bar{\psi}^1_i\psi^2_j+\bar{\psi}^2_i\psi^1_j)]}\ .
\eea

We perform a change of variables of Jacobian $-1$ to:
\bea
\chi^1_i=\frac{\psi^1_i+i\psi^2_i}{\sqrt{2}};
\chi^2_i=\frac{\psi^1_i-i\psi^2_i}{\sqrt{2}}\ .
\eea
As:
\bea
\bar{\chi}^1_i\chi^1_j=
\frac{1}{2}(\bar{\psi}^1_i\psi^1_j-i\bar{\psi}^2_i\psi^1_j+
i\bar{\psi}^1_i\psi^2_j+\bar{\psi}^2_i\psi^2_j),
\eea
we see that:
\bea
\Delta=(-1)^n \int\prod_k d\bar{\chi}^1_k d\chi^1_k
d\bar{\chi}^2_k d\chi^2_k
e^{-[a_{i j}(\bar{\chi}^1_i\chi^1_j+\bar{\chi}^2_i\chi^2_j)-
b_{i j}(\bar{\chi}^1_i\chi^1_j-\bar{\chi}^2_i\chi^2_j)]} \ .
\eea
Separating the terms in $\bar{\chi}^1\chi^1$ and $\bar{\chi}^2\chi^2$ proves (\ref{rdeterminant}).

The inverse matrix is divided into $2\times 2$ blocs with indices $ij$, according 
to the values $a,b=1,2$.
\bea
(R^{-1})^{ab}_{i j}=
\frac{\int d\bar{\psi}^1{d\psi}^1 d\bar{\psi}^2{d\psi}^2 \psi^{a}_{i}\bar{\psi^{b}_{j}}
e^{-\bar{\psi}A\psi}}
{\int d\bar{\psi}^1{d\psi}^1 d\bar{\psi}^2{d\psi}^2 e^{-\bar{\psi}A\psi}}\ .
\eea

The four elements of the blocs are 
given by (taking into account that the integral decouples so that all the crossed terms are zero):
\bea
\psi^{1}_{i}\bar{\psi}^1_{j}&=&
\frac{(\chi^{1}_{i}+\chi^2_{i})(\bar{\chi}^1_j+\bar{\chi}^2_j)}{2}=
\frac{1}{2}(\chi^{1}_{i}\bar{\chi}^1_j+\chi^2_{i}\bar{\chi}^2_j),
\nonumber\\
\psi^1_{i}\bar{\psi}^2_j&=&
\frac{(\chi^{1}_{i}+\chi^2_{i})(\bar{\chi}^1_j-\bar{\chi}^2_j)}{2(-i)}=
\frac{i}{2}(\chi^{1}_{i}\bar{\chi}^1_j-\chi^2_{i}\bar{\chi}^2_j),
\nonumber\\
\psi^2_i\bar{\psi}^1_j&=&\frac{(\chi^{1}_{i}-\chi^2_{i})(\bar{\chi}^1_j+\bar{\chi}^2_j)}{2i}=
\frac{-i}{2}(\chi^{1}_{i}\bar{\chi}^1_j-\chi^2_{i}\bar{\chi}^2_j),
\nonumber\\
\psi^2_i\bar{\psi}^2_j&=&
\frac{(\chi^{1}_{i}-\chi^2_{i})(\bar{\chi}^1_j-\bar{\chi}^2_j)}{2(-i)i}=
\frac{1}{2}(\chi^{1}_{i}\bar{\chi}^1_j+\chi^2_{i}\bar{\chi}^2_j).
\eea
(\ref{rinverse}) follows then easily.
\qed

Returning to our initial problem we remark that the matrix $A$
is the symmetric part coming from the propagator, and the oscillating part,
when symmetrized, leads naturally to an antisymmetric matrix $B$ times
the antisymmetric $\sigma_y$, so that in our case 
\bea
\det Q &=& [\det (A+B)(A-B)]^{D/2}\nonumber\\ 
&=& [\det (A+B)(A^t + B^t)]^{D/2} =  
[\det(A+B)]^D .
\eea

The propagator part is:
\bea \label{defmatrixa}
 A=\begin{pmatrix} S & 0 & 0\\ 0  & T & 0 \\ 0&0&0\\
\end{pmatrix}
\eea 
where $S$ and $T$ are the two diagonal $L$ by $L$ matrices
with diagonal elements $c_l = \coth(\frac{\alpha_l}{2}) = 1/t_l $, 
and $t_l = \tanh(\frac{\alpha_l}{2}) $, 
and the last lines and columns of zeroes reflect the purely
oscillating nature of the hypermomenta integrals.

The hypermomenta oscillations are (in the case of (\ref{hyperpnoncan})):
\bea
C_{v l}=\begin{pmatrix}
\sum_{i\in v}(-1)^{i+1}\epsilon_{l i} \\
\sum_{i\in v}(-1)^{i+1}\eta_{l i} \\
\end{pmatrix}\ .
\eea
Remark that the elements of $C$ are integers which can take only values $0$ or $ \pm 1$.
It is easy to check that for a connected graph $G$ the rank of the matrix $C$ is maximal, 
namely $n-1$. Picking a tree of $G$ proves that this is even true for 
the $L$ by $n$ lower part of $C$, corresponding to the long variables $v$ only.

To generalize to (\ref{hyperpcan}), we simply need to add another column to $C$, the one
corresponding to $p_{\bar v}$. The rank of the extended $2L$ by $n$ matrix $\bar C$
is then $n$, but the rank of the restriction of  $\bar C$ to its lower part 
corresponding to
the long variables $v$ is either $n-1$ or $n$ depending on whether the graph is orientable or not \cite{GMRV}.
This has important consequences for power counting.

The determinant of the quadratic form is the square of the determinant of the matrix $A + B$,
where
\bea\label{defmatrixb}
B= \begin{pmatrix}\frac{1}{4\theta\Omega} E & C \\
-C^t & 0 \\
\end{pmatrix}\ .
\eea
We can explicitate the oscillation part between the $u,v$ variables as  the $2L$ by $2L$ matrix E. This matrix  $E=\begin{pmatrix}E^{uu} & E^{uv} \\ E^{vu} & E^{vv} \\
\end{pmatrix}$  represents the vertices oscillations. One can check
\bea
E^{vv}_{l,l'}&=&\sum_v
\sum_{i\ne j ; \; i, j \in v}  (-1)^{i+j+1} \omega(i,j)\eta_{l i}\eta_{l' j},
\nonumber\\
E^{uu}_{l,l'}&=&\sum_v
\sum_{i\ne j ; \; i, j \in v}  (-1)^{i+j+1} \omega(i,j)\epsilon_{l i}\epsilon_{l' j},
\nonumber\\
E^{uv}_{l,l'}&=&\sum_v
\sum_{i\ne j ; \; i, j \in v}  (-1)^{i+j+1} \omega(i,j)\epsilon_{l i}\eta_{l' j},
\eea 
where we recall that $\omega(i,j) = 1$ if $i < j$ and  $\omega(i,j) = -1$ if $i > j$;
moreover $E^{vu}_{l,l'} = - E^{uv}_{l',l}$. Remark that
the matrix elements of $E$ are integers and can in fact
only take values $0, \pm 1, \pm 2$. Moreover $E_{l,l'}$ is zero
if $l$ and $l'$ do not hook to any common vertex; it can take value $\pm 2$ 
only if the two lines hook to at most two vertices in total, which is not generic, at least
for large graphs.

\begin{lemma}
Let $A=(a_i\delta_{i j})_{i,j\in\{1,\dotsc,N\}}$ 
be diagonal and $B=(b_{i j})_{i,j\in\{1,\dotsc,N\}}$ 
be such that $b_{ii}=0$ (we need not require $B$ antisymmetric). We have:
\be      
\det(A+B)=\sum_{K\subset \{1,\dotsc ,N\}}\det(B_{\hat{K}}) \prod_{i\in K}a_i 
\ee
where $B_{\hat{K}}$ is the matrix obtained from $B$ by deleting the lines and 
columns with indices in $K$.
\end{lemma}

\prf The proof is straightforward. We have:
\bea
&&\det(A+B)=\sum_{\sigma\in {\mathfrak S}_N}\epsilon_{\sigma}\prod_{i\in\{1,\dotsc,N\}}
(a_{i\sigma(i)}+b_{i\sigma(i)}) \nonumber\\
&&=\sum_{K\subset\{1,\dotsc,N\}}\prod_{i\in K}a_{i}
\sum_{\substack{\sigma\in {\mathfrak S}_N \\ \sigma(i)=i~\forall i\in K}}
\prod_{k\in \{1,\dotsc,N\}\setminus K}\epsilon(\sigma)b_{k\sigma(k)}
\eea
and the lemma follows.
\qed

In our case the matrix $B=\begin{pmatrix} \frac{1}{4\theta\Omega}E & C \\
-C^t & 0 \\ \end{pmatrix} $ is antisymmetric. 
Remark that the matrix 
\bea B'=\begin{pmatrix} E & C \\
-C^t & 0 \\ \end{pmatrix} 
\eea 
has integer coefficients.
Moreover $A$ in  (\ref{defmatrixa})  has zero diagonal
in the $n-1$ by $n-1$ lower right corner 
corresponding to hypermomenta. Exploiting these facts
we can develop $\det (A+B) $  into Pffafians to get:
\begin{lemma}
With $A$ and $B$ given by (\ref{defmatrixa}) and (\ref{defmatrixb})
\bea\label{pffafianfor}
\det (A+B) = \sum_{\substack{I\subset \{1\dotsc L\},  J\subset \{L+1\dotsc 2L\},\\ n + |I|+|J|\; {\rm odd}}}
 (4\theta\Omega)^{|I|+|J| + n-1-2L}   n^2_{I J}
\prod_{l\in I}c_l\prod_{l'\in J}t_{l'}
\eea
with $n_{I J}=\mathrm{Pf}(B'_{\hat{I}\hat{J}})$, the Pffafian of the oscillation matrix 
with deleted lines and columns $I$ among the first $L$ indices 
(corresponding to short variables $u$) and $J$ among the next $L$ 
indices (corresponding to long variables $v$). 
\end{lemma}
\prf
Since $A$ has the form  (\ref{defmatrixa}), the part $K$ of the previous lemma
has to be the disjoint union of two sets $I$ and $J$ respectively corresponding to
short and long variables. Once these sets are deleted from the matrix $B$
we obtain a matrix $B_{\hat{I}\hat{J}}$ which has size $2L -|I|-|J| +  n-1$. 
This matrix is antisymmetric, so its determinant is the square of the corresponding triangular Pfaffian.
The Pfaffian of such a matrix is zero unless its size $2L -|I|-|J| +  n-1 = 2p$ is even, in which case it is a sum, 
with signs, over the pairings of the $2p$ lines into $p$ pairs of the products of the corresponding
matrix elements. Now from the particular form of matrix $B$ which has a lower right block $0$, 
we know that any pairing
of the $n-1$ hypermomentum variables must be with an $u$ or $v$ variable. Hence
any pairing contributing to the Pfaffian has necessarily $n-1$ terms of the $C$ type, hence 
$[(2L -|I|-|J|)-(n-1)] /2$ terms of the $E/4\theta\Omega$ type, hence 
\bea \mathrm{Pf}(B_{\hat{I}\hat{J}}) = \frac{1}{(4\theta\Omega)^{L-(n+|I|+|J|-1)/2}} 
\mathrm{Pf}(B'_{\hat{I}\hat{J}})\ .
\eea
Therefore
\bea \det (B_{\hat{I}\hat{J}}) = \frac{1}{(4\theta\Omega)^{2L-n-|I|-|J|+1}} 
\mathrm{Pf}^2(B'_{\hat{I}\hat{J}}),
\eea
hence the Lemma holds, with $n_{I,J}=\mathrm{Pf}(B'_{\hat{I}\hat{J}})$ which must be
an integer since any Pfaffian with integer entries is integer.
\qed

We have thus expressed the determinant of
$Q$ as a sums of positive terms. 

Recalling that $\det Q = (\det (A+B) )^{D}$,
the amplitude ${\cal A}_{G,{\bar v}} (0)$ with external arguments $x_e$ and 
$p_{\bar v}$ put to 0
is nothing but (up to an inessential normalization)
\bea
{\cal A}_{G,{\bar v}} (0) = \int_{0}^{\infty} \det (A+B) ]^{-D/2}\prod_l 
\big[\frac{1-t_l^2}{t_l}\big]^{D/2} d\alpha_l  .
\eea
Putting $s=(4\theta\Omega)^{-1}$, we use the relation $2-2g = n-L+F$
and define the integer $k_{I,J} = \vert I\vert+\vert J\vert - L - F +1$ to get:
\bea\label{hypnoncanpfaff}
HU_{G,{\bar v}} (t) &=&  \sum_{I,J}  s^{2g-k_{I,J}} \ n_{I,J}^2
\prod_{l \not\in I} t_l \prod_{l' \in J} t_{l'}\ .
\eea
This is our precise definition of the normalization of the polynomial 
$HU_{G,{\bar v}}$ in the variables $t_l$ introduced in (\ref{hyperpnoncan}).
This normalization is adapted so that the limit $s \to 0$ will give back the ordinary Symanzik polynomial at leading order as $t$'s go to 0 (the ultraviolet limit), as shown in the next section.

To get the polynomial $HU_{G}$ in (\ref{hyperpcan}), we proceed exactly in the same
way, replacing $C$ by $\bar C$, and obtain that it is also
a polynomial in the variables $t_l$ with positive coefficients, which (up
to the factors in $4\theta\Omega$) are squares of integers. 
But we will see that the leading terms studied 
in the next section will be quite different in this case.

\subsection{Leading terms in the First Polynomial $HU$}

By leading terms, we mean terms which have the smallest global degree 
in the $t$ variables, since we are interested in power counting in the ``ultraviolet"
regime where all $t$'s are scaled to 0.
Such terms are obtained by taking $\vert I\vert $ maximal and $\vert J\vert $ minimal  in (\ref{pffafianfor}).
We shall compute the leading terms corresponding to $I=[1,...,L]$ hence taking all 
the $c_l$ elements of the diagonal 
and $J$ minimal so that the remaining minor is non zero
\footnote{These are not the only globally leading terms; there are terms whose global scaling is equivalent,
for example the $t^2_3$ term in (\ref{hueye1})-(\ref{hueye2}). By the positivity of $HU$ they can certainly not deteriorate the power counting established in this section, but only improve it in certain particular ``Hepp sectors".}.
Below we prove that such terms have $\vert J \vert =F-1$, 
This explains the normalization in (\ref{hypnoncanpfaff}).

\begin{figure}
\centerline{\epsfig{figure=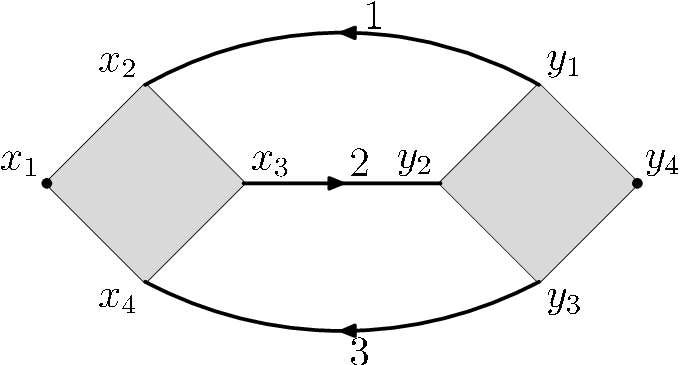,width=6cm} \hfil \epsfig{figure=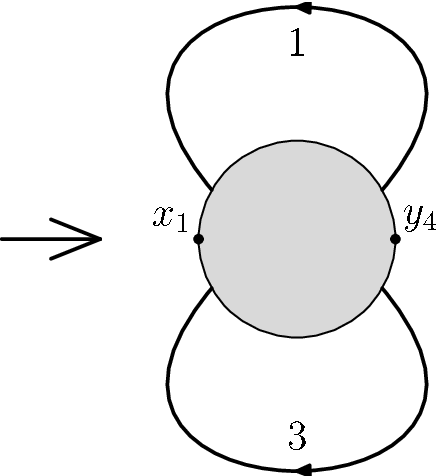,width=2cm}}
\caption{The First Filk Move on the Sunshine Graph}\label{firstfilk}
\end{figure}

\begin{figure}
\centerline{\epsfig{figure=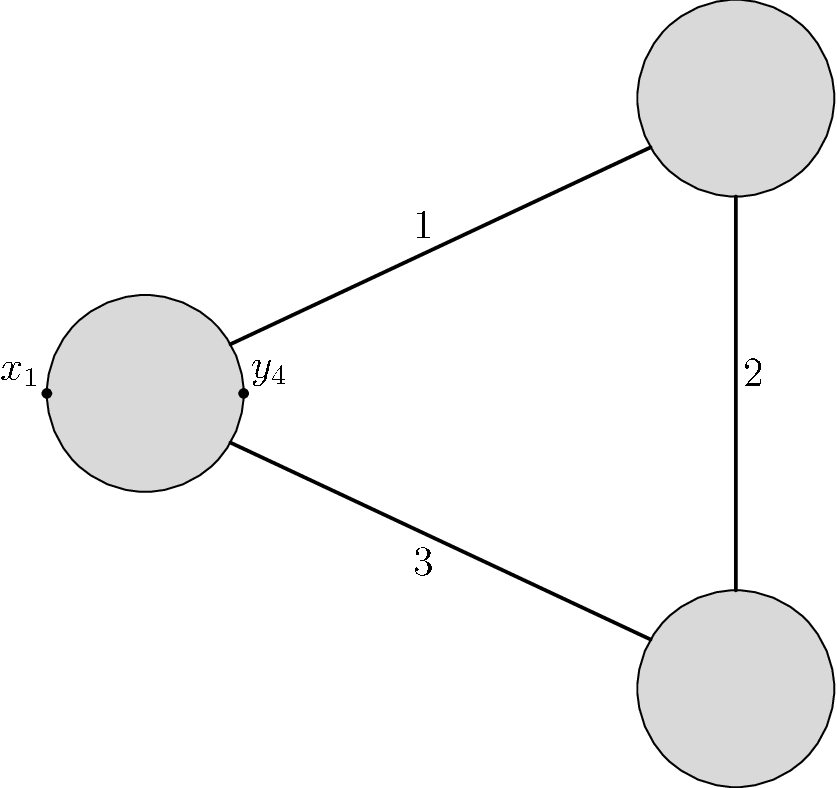,width=4cm} \hfil \epsfig{figure=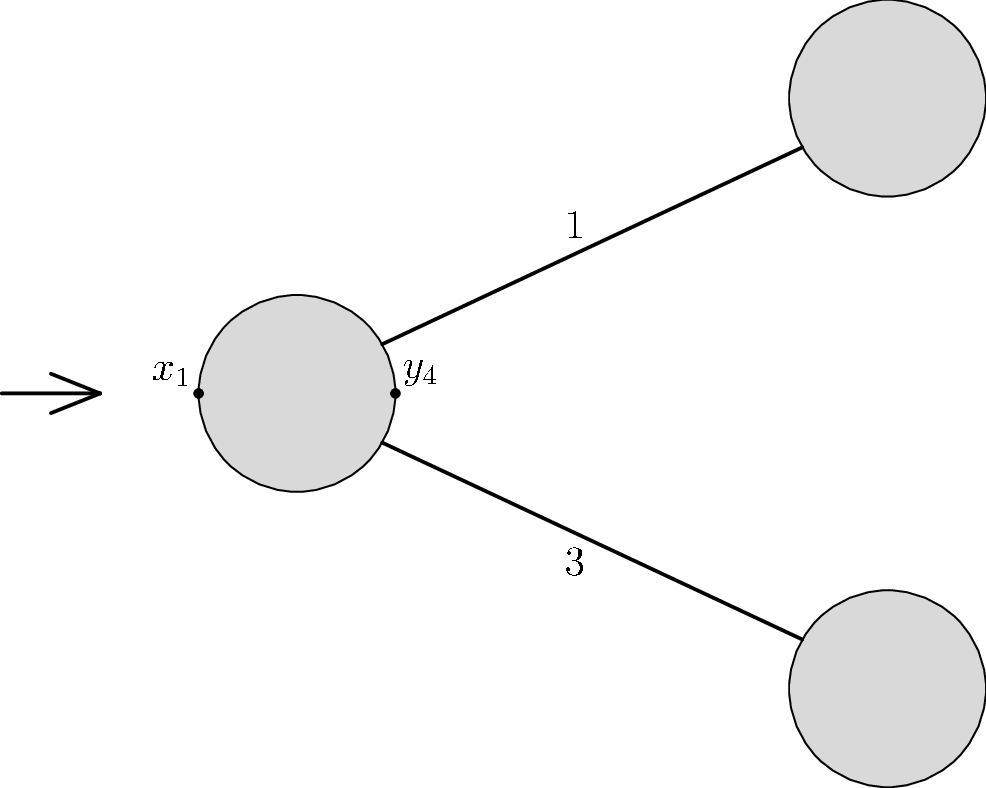,width=4cm}}
\caption{The First Filk Move on the dual of the Sunshine Graph}\label{dualsunshine}
\end{figure}

To analyze such leading terms we generalize the method of Filk's moves \cite{F},
defining three distinct topological operations on a ribbon graph.
The first one is a regular ``first Filk move", namely reduction of a tree line of the graph. This amounts to glue 
the two end vertices of the line (of coordination $p$ and $q$) to get a "fatter" vertex of coordination $p+q-2$. The new graph thus obtained has one vertex less and one line less. 
Since $2-2g=n-L+F$, this operation conserves the genus.
On Figure \ref{firstfilk} the contraction of the central line of the Sunshine Graph 
(also pictured on Figure \ref{figex2}) is shown.
In the dual graph this operation deletes the direct tree line, as shown on Figure \ref{dualsunshine}.

\begin{figure}
\centerline{\epsfig{figure=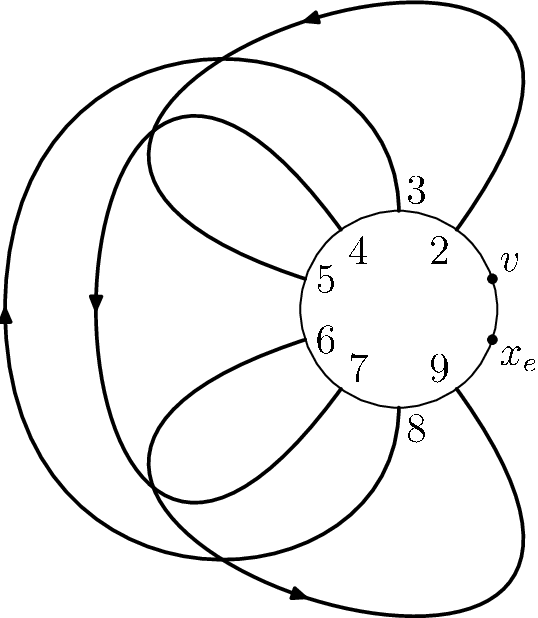,width=5cm}}
\caption{A Rooted Rosette}\label{rosette}
\end{figure}

Iterating this operation maximally we can always reduce a 
{\it spanning tree} in the direct graph,
with $n-1$ lines, obtaining a {\it rosette}. 
We recall that a rosette is simply a ribbon graph with a single vertex.
The rosettes we consider all have a root (i.e. an external line on $\bar v$), 
and a cyclic ordering to turn around, 
e.g. counterclockwise. We always draw the rooted rosette 
so that no line arches above the root. This defines uniquely 
a numbering of the halflines (see Figure \ref{rosette}, where the arrows represent the former line orientations\footnote{For an {\it orientable} graph these arrows
are compatible with the numbering, in the sense that halflines with even numbers
enter the rosette and halflines with odd numbers exit the rosette.}).

\begin{figure}
\centerline{\epsfig{figure=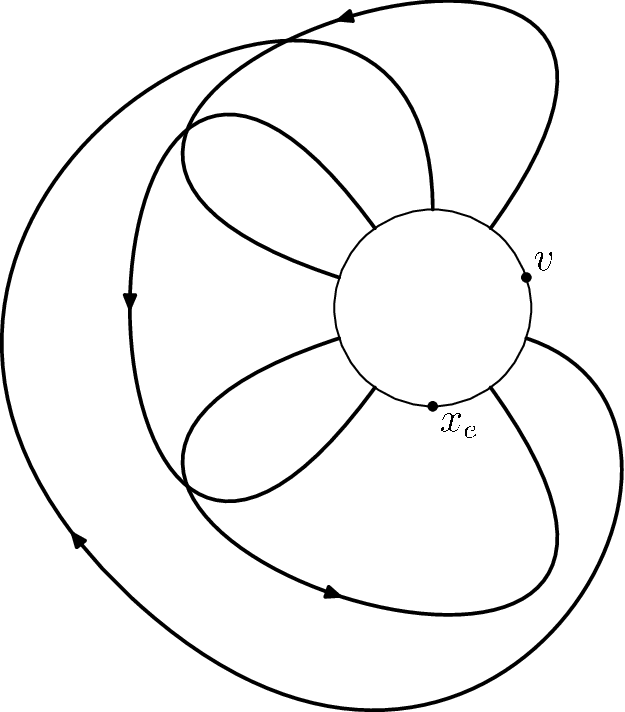,width=5cm}}
\caption{A SuperRosette}\label{superrosette}
\end{figure}

The second topological operation is the reduction of a tree line in the dual graph, exactly 
like the previous operation. Therefore it deletes this line in the direct graph. The resulting direct graph 
again keeps the same genus (remember that the genus of a graph is the same as the one of its dual). 
Iterating these two operations maximally we can always reduce completely a direct tree with $n-1$
lines {\it and} a dual tree with $f-1$ lines. We end up with a graph which we call a 
{\it superrosette}, which has only one vertex and one face 
(therefore its dual has one vertex and one face and is also a superrosette) (see Figure \ref{superrosette}).

The third operation is a genus reduction on a rosette. 
We define a {\it nice crossing} in a rosette to be a pair of lines such that the end point of the first is the successor in the rosette of the starting point of the other (in the natural cyclic order of the rosette). 
This ensures that the two lines have a common ``internal face".
When there are crossings in the rosette, it is easy to check that there exists at least one such nice crossing, for instance lines 2-5 and 4-8 in Figure \ref{rosette}.

\begin{figure}
\centerline{\epsfig{figure=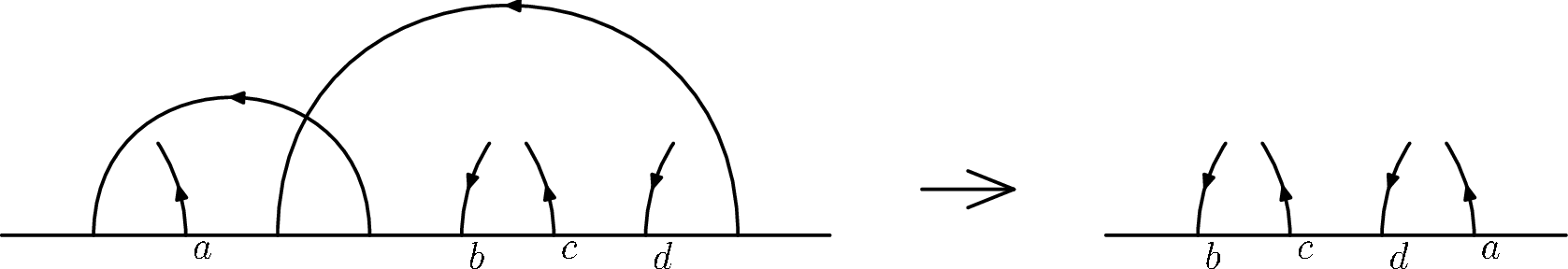,width=12cm,height=2cm}}
\caption{The Third Filk Move}\label{thirdfilk}
\end{figure}

The genus reduction consists in deleting the lines of a nice crossing
and interchanging all the halflines encompassed by the first line with those encompassed by the second line, see Figure \ref{thirdfilk}. 
This operation which we call the ``third Filk Move"\footnote{The second Filk move is 
the trivial simplification of non crossing lines in the rosette.} 
decreases the number of lines by two, glues again the faces in a coherent way, and decreases the genus by one.

We need then to compute the determinant of $B'$ matrices 
corresponding to reduced graphs of the type:
\bea
\begin{pmatrix}
\sum_{i\ne j} \omega(i,j) \eta_{l i}(-1)^{i+j+1}\eta_{l' j} 
& \sum_{i\in v} (-1)^{i+1}\eta_{l i} \\
\sum_{i\in v} (-1)^{i}\eta_{l i} & 0  \\
\end{pmatrix} \ .
\eea
As the graph is orientable and up to a possible overall sign we can cast the matrix into the form:
\bea
\begin{pmatrix}
\sum_{i\ne j}\omega(i,j)\epsilon_{l i}\epsilon_{l' j}
& \sum_{i\in v} \epsilon_{l i} \\
-\sum_{i\in v} \epsilon_{l i} & 0  \\
\end{pmatrix} \ .
\eea
We claim:
\begin{lemma}
\label{LemmaPfaff}
The above determinant is:
\begin{itemize}
\item  $0$ if the graph has more than one face,
\item $2^{2g}$ if the graph has exactly one face.
\end{itemize}
\end{lemma}

\prf  We reduce a tree. At each step we have a big vertex (the ``rosette in the making") $V$ and a small vertex $v$ bound to $V$  by a line $l=(i,j)$ which we contract. We then have at each step a Pfaffian 
$\int\prod_{d\chi_l d\psi_v}e^{-B'}$ where
\bea
B'&=&\sum_{l,l'}\chi_{l}\big(\sum_{i\ne j;i,j\in V}\omega(i,j)
\epsilon_{l i}\epsilon_{l' j} +
\sum_{i\ne j;i,j\in v}\omega(i,j)\epsilon_{l i}
\epsilon_{l' j}\big)\chi_{l'}\nonumber\\
&+&\sum_{l,v}\chi_l\epsilon_{l v}\psi_v\ .
\eea
At each step we use a permutation to put $l$ at the first place in the matrix. 
Note that this permutation has nothing to do with the ordering of the halflines. 
The terms containing $\chi_l$ or $\psi_v$ at each step are:
\bea
B'_l=&&\chi_l\epsilon_{l i}\sum_{l'}\big(\sum_{i\ne p;p  \in V}
\omega(i,p)\epsilon_{l' p} 
+\sum_{j\ne k;k  \in v}\omega(j,k)\epsilon_{l' k}\big)\chi_{l'}\nonumber\\
&&+\chi_l \epsilon_{l j}\psi_v-
\sum_{l''}\sum_{k\in v;k \neq j}\chi_{l''}\epsilon_{l'' k}\psi_v\ .
\eea

We perform the triangular change of variables:
\bea
\bar{\chi_l}&=&\chi_l-\epsilon_{l j}\sum_{l''}\sum_{k \in v;k\neq j}
\epsilon_{l'' k}\chi_{l''}\\
\bar{\psi_v}&=&\epsilon_{l j}\psi_v+\epsilon_{l i}\sum_{l'}
\big(\sum_{i\ne p;p \in V}\omega(i,p) \epsilon_{l' p}+
\sum_{i<k;k  \in v} \omega(i,k)\epsilon_{l' k}\big)\chi_{l'} 
\nonumber
\eea
Under this change of variable:
\bea
B'_l&=&\chi_l\bar{\psi_v}-\sum_{l''}\sum_{k\in v;k\neq j}\chi_{l''}
\epsilon_{l'' k} \psi_v\nonumber\\
&=&\bar{\chi_l}\bar{\psi_v}
-\epsilon_{l i}\epsilon_{l j}\sum_{l''}\sum_{l'}\chi_{l''}
\big(\sum_{k\in v;k \neq j}\sum_{p\in V;i\ne p}\omega(i,p)
\epsilon_{l'' k}\epsilon_{l' p}
\nonumber\\
&+&\sum_{k\in v;k\neq j}\sum_{r\in v;j\ne r}\omega(j,r)
\epsilon_{l'' k}\epsilon_{l' r}\big)\chi_{l'}
\eea
 
As $l$ is orientable $\epsilon_{l i}\epsilon_{l j}=-1$, 
so that the new term corresponds exactly to a new big vertex $\tilde{V}$ where the ordered halflines $k$ of the small vertex $v$ replace the halfline $i\in l$.   
We continue this procedure until we have reduced a complete tree in our graph. 
The remaining Pfaffian is of the form:
\bea
B'=\sum_{l<l';l\cap l'}\chi_l\big(\sum_{i\ne j;i,j\in V}\omega(i,j)\epsilon_{l i}\epsilon_{l'j}\big)\chi_{l'}
\eea
where $l<l'$ means that the starting halfline of $l$ precedes the end halfline of $l'$
in the final rosette vertex $R$.

When two lines in the rosette do not cross, in both sums over $i$ and $j$ above 
the two endpoints of any of the two lines add and give zero\footnote{This is the content of the ``second Filk move" of \cite{F}.}.
Consider now two lines $l_1=(i_1,j_1)$ and $l_2=(i_2,j_2)$ which cross each other, i.e. such that $i_1<i_2<j_1<j_2$. 
We have:
\bea
\chi_{l_1}(\epsilon_{l_1 i_1}\epsilon_{l_2 i_2}+\epsilon_{l_1 i_1}
\epsilon_{l_2 j_2}+
\epsilon_{l_1 j_1}\epsilon_{l_2 j_2}-\epsilon_{l_1 j_1}\epsilon_{l_2 i_2})
\chi_{l_2}=2\chi_{l_1}\epsilon_{l_1 i_1}\epsilon_{l_2 i_2}\chi_{l_2}.
\eea

Changing the variables $\chi_{l}\rightarrow\epsilon_{l i}\chi_l$
and writing $l\cap l'$ if $l$ crosses $l'$ we have:
\bea\label{prepthird}
 \frac{B'}{2}=\sum_{l<l';l\cap l'}\chi_{l}\chi_{l'}\ .
\eea

We perform a new simplification trick which we call the ``third Filk move".
We can pick two lines $l_1$ and $l_2$ which form a ``nice crossing", i.e. 
the start of $l_2$ immediately precedes the end of $l_1$ in the rosette.
We define the variables:
\bea
\bar\chi_{l_1}=\chi_{l_1}+\sum_{l'<l_2;l'\cap l_2}\chi_{l'}-\sum_{l''>l_2;l''
\cap l_2}\chi_{l''},\nonumber\\
\bar\chi_{l_2}=\chi_{l_2}-\sum_{l'<l_1;l'\cap l_1}\chi_{l'}+\sum_{l''>l_1;l''
\cap l_1}\chi_{l''}\; ,
\eea
and get:
\bea
\bar\chi_{l_1}\bar\chi_{l_2}&=&\chi_{l_1}\chi_{l_2}-\chi_{l_1}\sum_{l'<l_1;l'
\cap l_1}\chi_{l'}+
\chi_{l_1}\sum_{l''>l_1;l''\cap l_1}\chi_{l''}\nonumber\\
&&+\sum_{l'<l_2;l'\cap l_2}\chi_{l'}\chi_{l_2}-\sum_{l''>l_2;l''
\cap l_2}\chi_{l''}\chi_{l_2}\\
&& +(\sum_{l'<l_2;l'\cap l_2}\chi_{l'}-\sum_{l''>l_2;l''\cap l_2}\chi_{l''})
(-\sum_{l'<l_1;l'\cap l_1}\chi_{l'}+\sum_{l''>l_1;l''\cap l_1}\chi_{l''})\ .
\nonumber
\eea
Denoting $B'_{l_1 l_2}$ all the terms which contain either $l_1$ or $l_2$ in $B'$ and consistently denoting $l_p$ the lines which cross $l_1$ and $l_q$ those which cross $l_2$ we have:
\bea
 B'_{l_1 l_2}&=&\bar{\chi_{l_1}}\bar{\chi_{l_2}}+
 \sum_{l_q<l_2;l_q\cap l_2}\chi_{l_q}\sum_{l_p<l_1;l_p\cap l_1}\chi_{l_p}
 -\sum_{l_q>l_2;l_q\cap l_2}\chi_{l_q}\sum_{l_p<l_1;l_p\cap l_1}\chi_{l_p}
\nonumber\\
 &-&\sum_{l_q<l_2;l_q\cap l_2}\chi_{l_q}\sum_{l_p>l_1;l_p\cap l_1}\chi_{l_p}+
 \sum_{l_q>l_2;l_q\cap l_2}\chi_{l_q}\sum_{l_p>l_1;l_p\cap l_1}\chi_{l_p}\ .
\label{eq:terms}
\eea

Suppose that $l_q$ is the first line crossing $l_2$ and $l_p$ is the last line crossing $l_1$. Suppose moreover that $l_q<l_2$, $l_p<l_1$. When changing to the new variables we must add to the rosette factor a term of the first type in the above expression $\chi_{l_q}\chi_{l_p}$. What is the effect of such a term? 
If $l_p<l_q$ we have a crossing $\chi_{l_p}\chi_{l_q}$ and adding the above term gives zero. If $l_p>l_q$ adding the extra term acts like a new crossing. In both cases this amounts to permute the endpoints of $l_p$ and $l_q$. We can check that this holds
in fact in all cases, and that by induction the extra terms in equation
(\ref{eq:terms}) permute all the legs of $l_p$ type with all the legs
of $l_q$ type. We conclude that:
\bea
B'=\bar{\chi_{l_1}}\bar{\chi_{l_2}}+B'_{l_p\leftrightarrow l_q} \ ,
\eea
which is the content of our ``third Filk move".

With this change of variables in the Pfaffian we see that 
contracting the lines of the tree through first Filk moves, and 
reducing the genus on the rosette through third Filk moves, we end up
with a final Pfaffian which is $\pm 2^g$, taking into account the factors $2$
in (\ref{prepthird}), hence a determinant which is $2^{2g}$.

Suppose that the initial graph has $L$ propagators, $n$ vertices and $F$ faces.
We have $2-2g=n-L+F$. We can reduce $n-1$ lines of a tree and $g$ 
nice crossings. The remaining rosette will 
have $L_{\mathrm{rest}}=F-1$ propagators, with no crossing. 
The only way for the remaining Pfaffian not to be zero is to have  $F=1$. This 
completes the proof of Lemma \ref{LemmaPfaff}.
\qed

Returning to the initial problem, we know that in order 
for the final graph to have a single face we need to reduce a tree $\tilde{T}$ 
in the dual graph ${\cal G}$. So when $I$ is maximal, $J$ must contain a
tree in ${\cal G}$. But $J$ cannot be too big either, because the complement of $J$
must contain a tree in $G$ otherwise we cannot pair all hypermomenta variables.
We say that $J$ is {\it admissible} if 
\begin{itemize}
\item it contains a tree $\tilde T$ in the dual graph
\item its complement contains a tree $T$ in the direct graph
\item The rosette obtained by removing the lines of $J$ and contracting the lines of $T$
is a superrosette, hence has a single face. 
\end{itemize}

In particular if $J$ is admissible,  we have $F-1 \le \vert J \vert \le F-1 +2g$, and $k_j = \vert J \vert - F +1$
is even and obeys $0\le k_J \le 2g$.

We have altogether proved that:
\begin{lemma}
Suppose $I$ is maximal, i.e. contains all lines.
The integer $n_{I,J}$ in (\ref{hypnoncanpfaff}) is non zero 
if and only if $J$ is admissible, in which case $n_{I,J}^2=2^{2g- k_J}$
\end{lemma}

For $I$ maximal and $J$ admissible we have $k_{I,J}= k_J $.
Hence using positivity to keep the terms with maximal $I$ that we have identified, we get
\bea\label{firstbound}
HU_{G,{\bar v}} \ge  \sum_{J \ {\rm admissible}} (2s)^{2g- k_J}
\prod_{l\in J}t_l   \ .
\eea
>From this formula power counting follows easily  by 
finding the leading terms under scaling of all $t's$ to 0, which are the ones with minimal $J$.
They correspond to $J$'s which are trees in ${\cal G}$, hence which have $k_J = 0$. 
Keeping only these terms in (\ref{firstbound}) we have the weaker bound: 
\bea\label{secondbound}
HU_{G,{\bar v}} \ge (2s)^{2g}\sum_{J \ {\rm tree}\ \in\cal{G}}\prod_{l\in J }t_l\, .
\eea
At $D=4$, and for a graph of genus $g$ with $N$ external lines, using $2-2g =n+F-L$ and $L=2n-N/2$, we get from (\ref{secondbound}) at least 
a power counting in $\lambda^{2g + (N-4)/2}d\lambda$, so that
we recover the correct power counting
as function of the genus\footnote{We recall this power 
counting is understood easily in the
matrix base representation \cite{GW1}-\cite{GW2}-\cite{RVW}, but 
was not fully derived up to now in direct position
space \cite{GMRV} because the ``third Filk move"
was not performed there.}.

The commutative limit can be recovered easily
as the limit $s\to 0$, in which case only the terms with $k_J = 2g$ survive.
These $J$'s are exactly the complements of the trees in $G$ so that 
$HU_{G,{\bar v}}$ reduce to the usual commutative 
Symanzik polynomial $U_G$ in the limit $s  \to 0$. 

It is interesting to notice that since $s=1/4\theta\Omega$ this limit $s  \to 0$ seems 
to correspond to $\theta \to \infty$. But this is an artefact of our conventions and use of
the direct space representation. Indeed in the limit $\theta \to \infty$ the vertex in $x$-space
becomes the usual vertex in $p$-space of the commutative theory, whereas in the limit $\theta \to 0$ 
the vertex in $x$-space becomes the usual vertex in $x$-space of the commutative theory!
This explains this paradox (remark that the usual parametric representation 
(\ref{symanzik}) has $p$-space external variables).

Remark also that for planar graphs the complement of a tree in the dual graph is a tree
in the ordinary graph, hence (\ref{firstbound}) and (\ref{secondbound})
are identical in this case, as the sum over compatible $J$'s reduce
to a sum over trees of either $G$ or ${\cal G}$.

The canonical polynomial $HU_G$ can be analyzed in a similar way, but there we need
to take out one factor $c$, to pair to the additional hypermomentum in the Pfaffian,
so that $\vert I\vert $ is at most $L-1$; and the leading terms 
have one additional $t$ factor when compared to $HU_{G,{\bar v}}$
(see Section \ref{secexamp} for examples).

The power counting improvement in the number of faces broken by 
external fields is obtained after smearing external positions with
smooth test functions. For this improvement 
we analyze now the second hyperbolic polynomial.

\section{The Quadratic form in the External Positions, or Second Hyperbolic Polynomial}

We analyse only the $HV_{G,{\bar v}}$ polynomial, as the canonical $HV_{G}$ can be obtained easily afterwards.
The $P$ matrix in (\ref{defMPQ1})-(\ref{defMPQ2}) has elements:
\bea \label{formuleavecs}
P_{eu_l}&= -\sum_{i\neq e}\omega(i,e)\epsilon_{li}\frac{\sigma}{4\theta\Omega}\,,
\ 
P_{ev_l} &= -\sum_{i\neq e}\omega(i,e)\eta_{li}\frac{\sigma}{4\theta\Omega}\,,
\\
P_{ep_v}&=-\sum_{v;e\in v}(-1)^{e+1}\frac{\sigma}{\Omega}\,,
\
P_{\bar{p}u_l}&=\frac{\sigma}{\Omega}\,.
\eea

Note that all the elements of $P$ are multiples of $\sigma$. Upon transposition
of $P$ and multiplication we will recover a minus sign. Therefore the 
quadratic form in the external positions (and hypermomentum $\bar{p}$) in (\ref{defMPQ2}) is:
\bea
\begin{pmatrix}
x_e & \bar{p} \\
\end{pmatrix}
PQ^{-1}P^{t}
\begin{pmatrix} x_e \\ \bar{p}
\end{pmatrix}
&=&-x_{e_1}P_{e_1\tau}Q^{-1}_{\tau\tau'}P_{e_2\tau'}x_{e_2}-
x_{e_1}P_{e_1\tau}Q^{-1}_{\tau\tau'}P_{\bar{p}\tau'}\bar{p}\nonumber\\
&&-\bar{p}P_{\bar{p}\tau}Q^{-1}_{\tau\tau'}P_{e_2\tau'}x_{e_2}
-\bar{p}P_{\bar{p}\tau}Q^{-1}_{\tau\tau'}P_{\bar{p}\tau'}\bar{p}\,.
\eea

The inverse matrix of $Q$ being of the form
\bea\label{formulinvers}
 Q^{-1}_{\tau\tau'}=\frac{(A+B)^{-1}_{\tau\tau'}+(A-B)^{-1}_{\tau\tau'}}{2}\otimes I_D+
 \frac{(A+B)^{-1}-(A-B)^{-1}}{2}\otimes \sigma\, ,
\eea
we conclude that the quadratic form has a real part given by the $ I_D$ terms and an imaginary (oscillating) part 
given by the $\sigma$ terms:  the power counting we are looking for can be deduced solely from the former. 
To ease the writing we generically denote the set $x_e,\bar{p}$ by $x_e$. Also, let 
$\mathrm{Pf}(B_{\hat{K}\hat{\tau}})$ be the Pfaffian of the matrix obtained from $B$ by deleting 
the lines and columns in the set ${K,\tau}$, where again $K= I \cup J$
can be decomposed according to short and long variables. We have the analog of (\ref{pffafianfor}):
\bea
\frac{HV_{G,{\bar v}}}{HU_{G,{\bar v}}}(X_e)=
\frac{1}{\det (A+B)} \sum_K\prod_{i\in K}a_{ii}
\Big{[}\sum_{e_1}x_{e_1}\sum_{\tau\notin K}P_{e_1\tau}\epsilon_{K\tau}
\mathrm{Pf}(B_{\hat{K}\hat{\tau}})\Big{]}^2 \, .
\label{HVgv1}
\eea 

Multiplying by the product of $t$'s to compensate for the same
product in $HU_{G,{\bar v}}$ we get:
\begin{lemma}
\label{lemmapfaffsecond}
The real part $HV^R_{G,v}$ of the quadratic form in the external positions is:
\bea
\frac{HV^R_{G,{\bar v}}}{HU_{G,{\bar v}}}(x_e)=
\frac{1}{HU_{G,{\bar v}}} \sum_K\prod_{i\not\in K}t _{i}
\Big{[}\sum_{e_1}x_{e_1}\sum_{\tau\notin K}P_{e_1\tau}\epsilon_{K\tau}
\mathrm{Pf}(B_{\hat{K}\hat{\tau}})\Big{]}^2 \,.  
\label{HVgv}
\eea 
\end{lemma}

\prf:
We represent the matrix elements $Q^{-1}_{\tau\tau';\otimes K_d}$ by Grassmann integrals. As $(A+B)=(A-B)^t$ we have, for the first part of (\ref{formulinvers}):
\bea
Q^{-1}_{\tau\tau';\otimes I_D}&=&\frac{1}{2}[(A+B)^{-1}_{\tau\tau'}+(A-B)^{-1}_{\tau\tau'})\nonumber\\
&=&\frac{1}{2\det(A+B)}\int(d\bar{\psi}d\psi)[\psi_{\tau}\bar{\psi}_{\tau'}+
\psi_{\tau'}\bar{\psi}_{\tau}]e^{-\bar{\psi}(A+B)\psi} \,.
\eea

We perform the Pfaffian change of variables (of Jacobian $\imath$):
\bea
\psi_j=\frac{\chi_j+\imath \eta_j}{\sqrt{2}};~\bar{\psi}=\frac{\chi_j-\imath \eta_j}{\sqrt{2}}\,,
\eea
and we get (recalling that $d=2L+n-1$):
\bea
Q^{-1}_{\tau\tau';\otimes I_D}
&=&\frac{1}{2\det(A+B)}\int \prod_j(d\eta_j d\chi_j) \imath^{d}[\imath(\eta_{\tau}\chi_{\tau'}+\eta_{\tau'}
\chi_{\tau})]
\nonumber\\
&&e^{-\frac{1}{2}(\chi A \chi-\imath\eta A\chi+\chi B\chi-
\imath \eta B\chi+\imath\chi A\eta+\eta A \eta+\imath\chi B\eta+\eta B\eta)}\,.
\eea
As $B$ is antisymmetric the crossed terms in $B$ add to zero; as $A$ is diagonal the crossed terms 
are the only ones which survive. Reordering the measure and developping the exponential term 
in $A$ we get:
\bea
&&\frac{1}{2\det(A+B)}\int \prod_j(d\eta_j d\chi_j) \imath^{d}[-\imath(\chi_{\tau'}\eta_{\tau}+
\chi_{\tau}\eta_{\tau'})]
e^{-\frac{1}{2}(\chi B\chi+\eta B\eta)+\imath\eta A\chi}\nonumber\\
 &&=\frac{1}{2\det(A+B)}\int \prod d\eta \prod d\chi (-1)^{\frac{d^2}{2}}
(-\imath)(\chi_{\tau'}\eta_{\tau}+\chi_{\tau}\eta_{\tau'})\nonumber\\
&&\sum_{K}\prod_{i\in K}\imath^{|K|}a_{ii}\eta_i\chi_i
e^{-\frac{1}{2}(\chi B \chi+\eta B\eta)}\,.
\eea
Factorizing the sums over elements in $A$ and reordering the variables we finally get:
\bea
Q^{-1}_{\tau\tau;\otimes I_D}
&=&\frac{1}{2\det(A+B)}\sum_K\prod_{i\in K}a_{ii}
(-1)^{\frac{d^2}{2}}(-\imath)\imath^{|K|}(-1)^{\frac{|K|(|K|+1)}{2}}\nonumber\\
&&\int  d\eta \int d\chi \prod_{i\in K}\chi_i\prod_{i\in K}\eta_i
(\chi_{\tau'}\eta_{\tau}+\chi_{\tau}\eta_{\tau'})
 e^{-\frac{1}{2}(\chi B \chi+\eta B\eta)}\,.
\eea
Note that the last integrals are nonzero only if $d-|K|-1$ is even, which implies that the global sign in the above expression is always plus. The remaining Grassmann integrals can be expressed as:
\bea
\int \prod_{\alpha=1\dotsc d}d\chi_{\alpha}\prod_{i\in K}\chi_i\chi_{\tau}e^{-\frac{1}{2}\chi B\chi}=
\epsilon_{K\tau}\mathrm{Pf}(B_{\hat{K}\hat{\tau}})\,,
\eea
where $\epsilon_{K\tau}$ is the signature of the permutation 
\bea
1\dotsc d\rightarrow 1\dotsc\hat{i_1}\dotsc \hat{i_p}\dotsc \hat{i_{\tau}}\dotsc di_{\tau}i_p\dotsc i_1\,.
\eea

We have then:
\bea
 Q^{-1}_{\tau\tau';\otimes I_D}=\frac{1}{\det(A+B)}\sum_K\prod_{i\in K}a_{ii}\epsilon_{K\tau}
 \mathrm{Pf}(B_{\hat{K}\hat{\tau}})\epsilon_{K\tau'}\mathrm{Pf}(B_{\hat{K}\hat{\tau'}})\,,
\eea
and the real part of the quadratic form is:
\bea
&& \sum_{e_1, e_2, \tau , \tau'}  -x_{e_1}P_{e_1\tau} \frac{1}{\det(A+B)}\sum_K\prod_{i\in K}a_{ii}
\epsilon_{K\tau}\mathrm{Pf}(B_{\hat{K}\hat{\tau}})
 \epsilon_{K\tau'}\mathrm{Pf}(B_{\hat{K}\hat{\tau'}})
P_{e_2\tau'}x_{e_2}\nonumber\\
&&=-\frac{1}{\det(A+B)}\sum_K\prod_{i\in K}a_{ii}
\Big{[}\sum_{e_1}x_{e_1}\sum_{\tau\notin K}P_{e_1\tau}\epsilon_{K\tau}\mathrm{Pf}(B_{\hat{K}\hat{\tau}})
\Big{]}^2\,.
\eea
This ends the proof of (\ref{HVgv}). \qed

Using similar methods one can prove that the imaginary part of the inverse matrix elements is:
\bea
Q^{-1}_{\tau\tau';\otimes\sigma}=\frac{1}{\det(A+B)}\sum_K\prod_{i\in K}a_{ii}
\epsilon_{K\tau\tau'}\mathrm{Pf}(B_{\hat{K}\hat{\tau}\hat{\tau'}})\epsilon_K\mathrm{Pf}(B_{\hat{K}})\,,
\eea
and consequently the contribution to the quadratic form is:
\bea
&&-\frac{1}{\det(A+B)}\sum_K\prod_{i\in K}a_{ii}\epsilon_K\mathrm{Pf}(B_{\hat{K}})
\nonumber\\
&&\Big{[}\sum_{e_1,e_2} \Big{(}
\sum_{\tau\tau'}P_{e_1\tau}\epsilon_{K\tau\tau'}\mathrm{Pf}(B_{\hat{K}\hat{\tau}\hat{\tau'}})
P_{e_2\tau'}\Big{)}
x_{e_1}\sigma x_{e_2}\Big{]}\,.
\eea

\subsection{Leading terms}

The last step of our analysis is to find the leading terms in eq. (\ref{HVgv}). In order to do this one must find under which conditions Pfaffians like:
\bea
\epsilon_{K\tau}\mathrm{Pf}(B_{\hat{K}\hat{\tau}})=
\int \prod_{\alpha=1\dotsc d}d\chi_{\alpha}\prod_{i\in K}\chi_i\chi_{\tau}e^{-\frac{1}{2}\chi B\chi}\,,
\eea
are nonzero.

A priori one can exploit the $\delta$ functions associated with each external vertex to simplify the oscillating factor involving the external position. These manipulations do not have any effect on the form of $HV$ but can be used to set some $P_{ev_l}$ 
and $P_{eu_l}$ to zero. This is always the case if our graph does not have any vertex with two opposite external points.
We conclude that the power counting behavior of the second polynomial should entirely 
be given by $P_{ev_e}$ and 
consequently by terms like:
\bea
\epsilon_{Kv_e}\mathrm{Pf}(B_{\hat{K}\hat{v_e}})=
\int \prod_{\alpha=1\dotsc d}d\chi_{\alpha}\prod_{i\in K}\chi_i\psi_{v_e}e^{-\frac{1}{2}\chi B\chi}\,.
\eea

The reasoning is similar to that we used to find the leading contributions in the $HU_{G,v}$. We must find a nonzero Pfaffian multiplied by all the $c$'s and the smallest number of $t$'s possible, hence corresponding to
$K= I \cup J$ with $I=[1,...L]$ maximal and $J$ minimal.
One could find a change of variables similar to those of lemma \ref{LemmaPfaff}, but we will use here a slightly different approach.
 
We introduce a dummy Grassmann variable $\Psi$ in the Pfaffian integrals, to have:
\bea
\epsilon_{Kv_e}\mathrm{Pf}(B_{\hat{K}\hat{v_e}})=
\int \prod_{\alpha=1\dotsc d}d\chi_{\alpha}d\Psi\ \Psi \prod_{i\in K}\chi_i\psi_{v_e}e^{-\frac{1}{2}\chi B\chi}\,.
\eea

Next we exponentiate $\Psi\psi_{v_e}$ and pass to the Pfaffian of a modified matrix $B'$ (corresponding to a modified graph $G'$). The modified graph is obtained from $G$ by adding a line $l_0$ from $x_e$ to the root external line. Moreover, following the reasoning of lemma \ref{LemmaPfaff} we see that our Pfaffian is not modified if we impose that the dummy line is constrained to be a tree line in the direct graph $G'$, i.e. has to pair with
an hypermomentum variable.
We then have a one to one correspondence between the leading term in $HV_{G,{\bar v}}$ and the leading terms for the first polynomial of the modified graph $HU_{G',{\bar v}}$ 
in which the dummy line $l_0$ is chosen as a tree line.

In order to conclude we need only to compute the genus of the modified graph $G'$.  
Suppose the external point $x_e$ broke another face than the root external point. 
The dummy line we added will identify the two faces so that $G'$ has
$n'=n$, $L'=L+1$, $F'=F-1$ and
$n'-L'+F'=2-2g'$, so that $g'=g+1$.
If, on the other hand, $x_e$ broke the same face as the root halfline, the dummy line will part the latter in two different faces. 
We then have $F'=F+1$ and consequently $g'=g$.

We note that to any tree in $G'$ constrained to contain the dummy line $l_0$ there corresponds 
a two-tree $T_2$ in $G$, that is a tree minus a line (by removing this dummy line).

Therefore by Lemma III.4 the Pfaffian we are considering is non zero only if the new reduced rosette, 
where the dummy line is contracted as a tree line, has exactly one face and $2g'$ lines. 

We obtain therefore the real part of $V_{G,{\bar v}}$ as a sum of 
positive terms and exact 
analogs of bounds (\ref{firstbound}) and (\ref{secondbound}). 
We say that $J$ is 2-admissible in $G$ if $J$ is admissible in $G'$ and
the dummy line $l_0$ is in a tree of $G'$ contained in the complement of $J$.

Let $\tilde{G}$, be the graph obtained from $G$ by 
deleting the lines in $J$ and contracting the two-tree $T_2$. It has two faces, the one broken by the root
 and another one, $F_J$. This $F_J$ contains typically the external points belonging to several broken faces 
 in the initial graph.
The dummy line $l_0$ will link this two faces, but the topological structure of $G'$ is the same no matter 
which external points are chosen in $F_J$. We will therefore obtain a sum over all this possible choices.
We conclude that the bound analog to (\ref{firstbound}) is
\bea\label{firstboundext}
HV^R_{G,{\bar v}}(x_e) \ge \sum_{ J \ {\rm 2-admissible \ in} \ G} 
(2s)^{2g'-k_J}\prod_{l\in J}t_l [ \sum_{e \in F_J  } (-1)^e x_e ]^2 \; .
\eea
 
The analog of (\ref{secondbound}) is
\bea\label{secondboundext}
HV^R_{G,{\bar v}}(x_e) \ge \sum_{ J \ {\rm tree\  in} \ {\cal G}'} 
(2s)^{2g'}\prod_{l\in J} t_l \  [\sum_{e \in F_J  } (-1)^e x_e ]^2 \; .
\eea
This bound is the one useful to extract
the power counting in the broken faces.
Indeed when integrating the remaining external variables
against fixed test functions and scaling each $t$ variable by $\lambda \to 0$, 
we recover the full exact power counting of $G$ namely in dimension $D=4$ the scaling
$\lambda^{2g +(B-1) + (N-4)/2}d\lambda$.
Indeed each broken face leads to a term in $e^{-  [x_e+\dotsc]^2}$ for some external variable $x_e$ of the broken face, hence to an improved factor $\lambda$ when integrated against a fixed test function. 

It is also possible to recover the second Symanzik polynomial as $s \to 0$ in 
(\ref{firstboundext}); indeed for $k_J = 2g'$, we find that $J$ is the complement of a tree in $G'$
containing $l_0$, hence of a two-tree in $G$. Moreover, in this case the face $F_J$ will be the external face 
of the conected component $E(T_{2})\in T_2$ not containing the root, and we recover the known invariant 
$(\sum_{e\in E(T_2)}(-1)^{e+1}x_e)^2$.

It remains to discuss the case with special ``diagonal" vertices, that is with opposite 
external arguments as in Figure \ref{figex4}.
In this case bounds (\ref{firstboundext}) and (\ref{secondboundext}) still hold because
although there is a sum of two Paffians in Lemma \ref{lemmapfaffsecond}, 
they cannot add up to 0; they correspond to graphs $G'_1$ (of the same kind as before) and 
$G'_2$ (of a new type, with one line erased) with genuses
$g'_1$ and $g'_2 =g'_1 - 1$. These Pfafians have the same scaling in $s$ because 
there is an additional $s$ for $G'_2$ coming from formula (\ref{formuleavecs}), but they have not the same
power of 2 hence their sum cannot be zero again!

There is another modification: the alternate sum over a face no longer appear
in (\ref{firstboundext}) and  in  (\ref{secondboundext})  
if the root vertex itself is of this diagonal type which is the case
for the ``Broken Bubble" graph of Figure \ref{figex4}. These modifications 
do not affect the power counting and their verification is left to the reader.

\section{Examples}
\label{secexamp}

In this section we give the exact expressions for several of our polynomials. We recall that 
$s =(4\theta\Omega)^{-1}$.

\begin{figure}[ht]
\centerline{\epsfig{figure=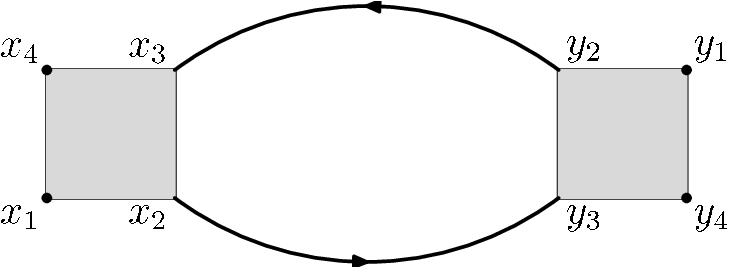,width=6cm}}
\caption{The Bubble graph}\label{figex1}
\end{figure}

We start by the bubble graph, Figure \ref{figex1}:

\bea
HU_{G,v}&=&(1+4s^2)(t_1+t_2+t_1^2t_2+t_1t_2^2)\,,\nonumber\\
HV_{G,v}&=&t_2^2\Big{[}p_2+2s(x_4-x_1)\Big{]}^2+t_1t_2\Big{[}2p_2^2+(1+16s^4)(x_1-x_4)^2
                \Big{]}\,,\nonumber\\
                &&+t_1^2\Big{[}p_2+2s(x_1-x_4)\Big{]}^2\nonumber\\
HU_{G}&=&4(t_1+t_2)^2\,,\nonumber\\
HV_{G}&=&(1+4s^2)\Big{[}(x_1-x_4+y_1-y_4)^2(t_1+t_2)\nonumber\\
                   && +(x_1-x_4-y_1+y_4)^2(t_1t_2^2+t_2t_1^2)\Big{]}\,.
\eea

The first two terms in $HU_{G,v}$ are the leading ones we previously exhibited. For the $HV_{G,v}$ we see that the scaling of the quadratic form will be in $O(1)$, which was expected as we do have only one broken face. Furthermore $HU_{G}$ scales in $t^2$, as expected, and the first term in the quadratic form $HU_{G}/HV_{G}$ is exactly the required one to reconstitute the $\delta$ function on the external legs in the $UV$ region.
The term $t_1t_2(x_1-x_4)^2$ in  $HV_{G,\bar{v}}$ is one of the leading terms previousely computed. It comes from the graphs $G'$ in which either $x_1$ or $x_4$ are linked to the root by the dummy line, and both lines $1$ and $2$ are chosen in the set $J$ in $\cal{G'}$.

Then comes the sunshine graph Fig. \ref{figex2}:
\begin{figure}[h]
\centerline{\epsfig{figure=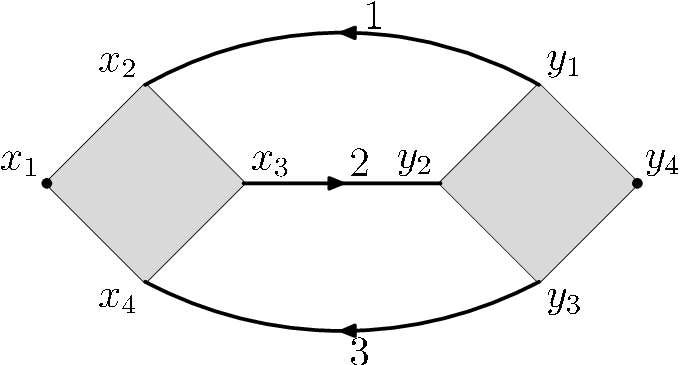,width=6cm}}
\caption{The Sunshine graph}\label{figex2}
\end{figure}

\bea
HU_{G,v}&=&\Big{[} t_1t_2+t_1t_3+t_2t_3+t_1^2t_2t_3+t_1t_2^2t_3+t_1t_2t_3^2\Big{]}
(1+8s^2+16s^4)\nonumber\\
&&+16s^2(t_2^2+t_1^2t_3^2)\, ,\nonumber\\
HU_{G}&=&t_1t_2t_3(1+64s^4)\nonumber\\
&&+\Big{[}t_1^2t_2+t_1t_2^2+t_1^2t_3+t_2^2t_3+t_1t_3^2+t_2t_3^2\Big{]}(4+16s^2)\,.
\eea

Here we identify also the leading contributions in $HU_{G,v}$, and the extra scaling factor in the $HU_{G}$.

For the nonplanar sunshine graph (see Fig. \ref{figex3}) we have:
\begin{figure}[ht]
\centerline{\epsfig{figure=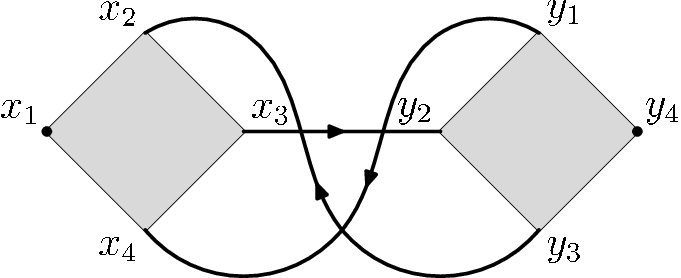,width=6cm}}
\caption{The Non-planar Sunshine graph}\label{figex3}
\end{figure}
\bea
HU_{G,v}&=&\Big{[} t_1t_2+t_1t_3+t_2t_3+t_1^2t_2t_3+t_1t_2^2t_3+t_1t_2t_3^2\Big{]}
(1+8s^2+16s^4)\nonumber\\
&&+4s^2\Big{[}1+t_1^2+t_2^2+t_1^2t_2^2+t_3^2+t_1^2t_3^2+t_2^2t_3^2+
t_1^2t_2^2t_3^2\Big{]}\,,\nonumber\\
HU_{G}&=&4\Big{[}t_1^2(t_2+t_3)+t_2t_3(t_2+t_3)+t_1(t_2^2+3t_2t_3+t_3^2)\Big{]}\nonumber\\
&&+16s^2\Big{[}t_1+t_2+t_3+t_1t_2^2t_3^2+t_1^2t_2t_3(t_2+t_3)\Big{]}
+64t_1t_2t_3s^4\,.
\eea

We note the improvement in the genus, as both $HU_{G,v}$ and $HU_{G,v}$ scale in $t^{-2}$ with respect to there planar counterparts.

For the broken bubble graph (see Fig. \ref{figex4}) we have:
\begin{figure}[ht]
\centerline{\epsfig{figure=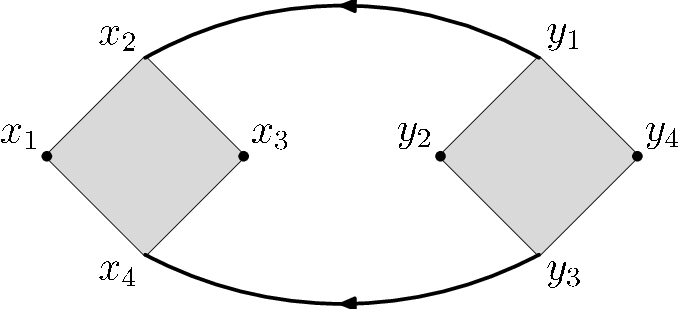,width=6cm}}
\caption{The Broken Bubble Graph}\label{figex4}
\end{figure}

\bea
HU_{G,v}&=&(1+4s^2)(t_1+t_2+t_1^2t_2+t_1t_2^2)\,,\nonumber\\
HV_{G,v}&=& t_2^2 \Big{[}4s^2(x_1+y_2)^2+(p_2-2s(x_3+y_4))^2\Big{]}+t_1^2\Big{[}p_2
+2s(x_3-y_4) \Big{]}^2\,,\nonumber\\
&&+t_1t_2\Big{[}8s^2y_2^2+2(p_2-2sy_4)^2+(x_1+x_3)^2+16s^4(x_1-x_3)^2\Big{]}\nonumber\\
&&+t_1^2t_2^24s^2(x_1-y_2)^2\,,\nonumber\\
HU_{G}&=&4(t_1+t_2)^2\,,\\
HV_{G}&=&(t_1+t_2)\Big{[}(y_2+y_4-x_1-x_3)^2+4s^2(y_2-y_4-x_1-3x_3)^2\Big{]}\nonumber\\
&&+(t_1^2t_2+t_1t_2^2)\Big{[}(y_2+y_4+x_1+x_3)^2+4s^2(y_2-y_4+x_1+3x_3)^2\Big{]}\,,\nonumber
\eea

Note that $HU_{G,v}$ and $HU_{G}$ are identical with those of the bubble with only one broken face. It is natural, as the amelioration in the broken faces for a given graph can be seen only in the $HV_{G,v}$ and $HV_{G}$.  Take $HV_{G}$. We see that we have two linear combinations which in the $UV$ region become an approximate $\delta$ function, whereas in the bubble graph with one broken face we had only one, therefore giving us the improvement in the broken faces. Similarely for the $HV_{G,v}$ we see that for the broken bubble we have three independent linear combinations which scale in $t^{-1}$ whereas for the bubble with only one broken face we only had two.
The term $t_1t_2(x_1+x_3)^2$ in $HV_{G,\bar{v}}$ is one of those computed in the previous section. Even if the two external points $x_1$ and $x_3$ do not belong to the same face they still appear summed, as we can chose the root halfline to be either
$y_2$ and $y_4$ and we must add the contributions for each choice. This is an example of the slight modifications generated by the presence of "diagonal" vertices.

\begin{figure}[ht]
\centerline{\epsfig{figure=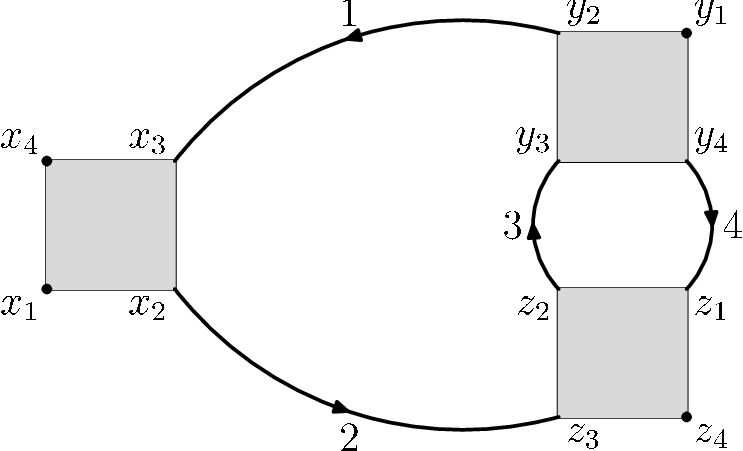,width=6cm}}
\caption{The Half-Eye Graph}\label{figeye}
\end{figure}
Finally, for the half-eye graph (see Fig. \ref{figeye}), we start by defining:
\bea
A_{24}=t_1t_3+t_1t_3t_2^2+t_1t_3t_4^2+t_1t_3t_2^2t_4^2\,.
\eea
The $HU_{G,v}$ polynomial with fixed hypermomentum corresponding to the vertex with two external legs is:
\bea\label{hueye1}
HU_{G,v_1}&=&(A_{24}+A_{14}+A_{23}+A_{13}+A_{12})(1+8s^2+16s^4)\nonumber\\
&&+t_1t_2t_3t_4(8+16s^2+256s^4)+4t_1t_2t_3^2+4t_1t_2t_4^2\nonumber\\
&&+16s^2(t_3^2+t_2^2t_4^2+t_1^2t_4^2+t_1^2t_2^2t_3^2)\nonumber\\
&&+64s^4(t_1t_2t_3^2+t_1t_2t_4^2)\,,
\eea
whereas with another fixed hypermomentum we get:
\bea\label{hueye2}
HU_{G,v_2}&=&(A_{24}+A_{14}+A_{23}+A_{13}+A_{12})(1+8s^2+16s^4)\nonumber\\
&&+t_1t_2t_3t_4(4+32s^2+64s^4)+32s^2t_1t_2t_3^2+32s^2t_1t_2t_4^2\nonumber\\
&&+16s^2(t_3^2+t_1^2t_4^2+t_2^2t_4^2+t_1^2t_2^3t_3^2)\,.   
\eea

Note that the leading terms are identical, the choice of the root perturbing only the non-leading ones. Moreover note the presence of the $t_3^2$ term. Its presence can be understood by the fact that in the sector $t_1,t_2,t_4>t_3$ the subgraph formed by the lines $1,2,4$ has two broken faces. This is the sign of 
a power counting improvement due to the additional broken face in that sector. 
To exploit it, we have just to integrate over the variables of line $3$
in that sector, using the second polynomial $HV_{G',v}$ for the triangle subgraph $G'$ 
made of lines $1,2,4$.  

Finally the canonical $HU_{G}$ polynomial is:
\bea
HU_{G}&=&(4+16s^2)(t_1^2t_3+t_1t_3^2+t_2^2t_3+t_2t_3^2+t_1^2t_4+t_1t_4^2+
t_2^2t_4+t_2t_4^2\nonumber\\
&&+t_3^2t_4+t_3t_4^2
+t_1^2t_2t_3^2+t_1t_2^2t_3^2+t_1^2t_2t_4^2+t_1t_2^2t_4^2
+t_1^2t_2^2t_3^2t_4\nonumber\\
&&+t_1^2t_2^2t_3t_4^2)
+(8+32s^2)(t_1t_2t_3+t_1t_2t_4+t_1t_2t_3^2t_4+t_1t_2t_3t_4^2)\nonumber\\
&&+(12+64s^4)
(t_1t_3t_4+t_2t_3t_4+t_1^2t_2t_3t_4+t_1t_2^2t_3t_4)\,.
\eea

\subsubsection*{Acknowledgment}
We are indebted to A.~Abdesselam for his inspiring reference \cite{A}
and his help on the track to the Pfaffian analysis of Sections III-IV. 
We also thank  M.~Disertori, J.~Magnen and F.~Vignes-Tourneret 
for useful discussions during preparation of this work.

\end{document}